\documentstyle[12pt,epsf]{article}
\title{$B_s (B_d) \to \gamma \nu \bar \nu $ }
\author{ Cai-Dian  L\"{u} and Da-Xin Zhang\\
 Physics Department, Technion- Israel Institute of Technology, \\
  Haifa 32000, Israel.}

\date{}
\begin{document}
\maketitle
\begin{picture}(0,0)(0,0)
\put(340,270){{\large hep-ph/9604378}}
\put(340,250){{\large TECHNION-PH-96-6}}
\put(340,230){{\large February, 1996}}
\end{picture}

\begin{abstract}
We study the loop induced rare decays $B_s(B_d)\to\gamma\nu\bar\nu$ 
within the standard model. Both constituent 
quark model and pole model are used and the branching ratios
 turn out to be of the orders of $10^{-8}$ for  $B_s\to \gamma 
\nu \bar \nu $ and  $10^{-9}$ for $B_d \to \gamma \nu \bar \nu $.
These processes can be served to determine the decay constants of 
$B_s$ and $B_d$.
\end{abstract}
\newpage

\section{Introduction}
Rare decays of bottom quark \cite{smrev} can be served 
as tests of the standard model (SM). As the observed process $b\to 
s\gamma$ which induced  $B\to K^*\gamma$ \cite{bkstar} 
and $B\to X_s\gamma$ \cite{bxs} at the hadronic level,
$b\to s l^+l^-$\cite{bsll} are also studied extensively in the past
both at inclusive and at exclusive levels.
It was recognized 
that the SM prediction for the branching ratio of $B\to X_s\nu\bar\nu$, 
which is $\sim 5\times 10^{-5}$\cite{buras}, 
is within one order beyond the present detectivity and this process
might be observed soon\cite{grossman}.
Furthermore, the corresponding exclusive decay $B\to K^*\nu \bar \nu$ 
is important as an input to  determine $V_{ub}$ to an accuracy of $10\%$
\cite{vub}.
 
Pure leptonic decays of heavy pseudoscalar mesons
into light lepton pairs are helicity suppressed,
their branching ratios are \cite{cam}:
\begin{eqnarray}
B(B_s\to\mu^+\mu^-)&=&1.8\times 10^{-9},\nonumber\\
 B(B_s\to e^+e^-)&=&4.2\times 10^{-14},
\end{eqnarray}
which make it difficult to  determine  $f_{B_s}$ from these processes.
For $B_d$ the situation is even worse due to the small CKM mixing angles.
Although the processes $B_s(B_d)\to\tau^+\tau^-$ do not suffer
from this suppression mechanism and the branching ratio is about
$10^{-7}$ \cite{buras} in SM,
 it is hard to be detected at future $B$-factory where
the efficiency is not better than $10^{-2}$.
The processes $B_s(B_d)\to\gamma\nu\bar\nu$, whose branching ratios
depend quadrically on $f_{B_s}(f_{B})$, can be taken as a possible
alternate to determine the decay constants.
For  detections of  these processes the method of searching for
the ``missing mass'' can be used, which is similar to the case of 
$B\to X_s\nu\bar\nu$ \cite{grossman}.

In the present work, we study the processes $B_s(B_d)\to\gamma\nu\bar\nu$
within the SM.
We will analyze  $B_s(B_d)\to\gamma\nu\bar\nu$ in the following:
In section 2 the relevant effective Hamiltonian will be given in SM.
Constituent quark model and pole model will be used in
section 3 to give the predictions.
Finally,  section 4 contains some brief discussion.

\section{Effective Hamiltonian}

Let us start with the quark level process $b  \to q \nu \bar \nu$, 
with $q=s$ or $d$.
The Feynman diagrams are displayed in Fig.1. Both  box and Z penguin 
diagrams contribute to this process. The resulting effective Hamiltonian
is left handed in SM \cite{lim,buras}:
\begin{equation}
{\cal H} = C (\bar q\gamma _\mu P_L b )(\bar \nu \gamma ^\mu P_L\nu),
\end{equation}
with $P_L=(1-\gamma_5)/2$.
The  coefficient $C$ is 
\begin{equation}
C=\frac{\sqrt{2} G_F  \alpha }{\pi \sin ^2 \theta _w} V_{tb} V_{tq}^* 
\frac{x}{8} \left( \frac{x+2}{x-1} +\frac{ 3x-6}{(x-1)^2} \ln x \right),
\label{c}
\end{equation}
where $x=m_t^2/m_W^2$. For simplicity, we neglect the QCD correction
to this coefficient, whose effects are within $2\%$ if
appropriate renormalization point is chosen \cite{buras}.

Since the neutrinos are massless in minimal SM, the processes
 $ B_q (b \bar q )\to \nu \bar \nu$
\footnote{We denote mesons with the quark content $(b \bar q )$  
as $ B_q$ for convenience. }
are forbidden by helicity. If an additional 
photon line is attached to any of the charged lines in Fig.1, the 
situation will be different: no helicity suppression exists any more. 
However, when the photon line is attached
to one of the internal charged lines, there will be a suppression factor 
of $m_b^2/m_W^2$ 
 in the Wilson coefficient compared with the ones for $b 
\to q \nu \bar \nu$. The reason is that these effective 
operators are now  dimension-8 instead of dimension-6. 
On the other 
hand, the diagrams in Fig. 1 with 
 photon line  connected to one of the external (bottom or strange 
quark) lines, whose effective Hamiltonian turn out to be
\begin{equation}
{\cal H} = -\frac{e}{6} C ~\bar q \left [ \not \! \epsilon_\gamma 
\frac{ \not \!p
_\gamma -\not \! p_q +m_q}{(p_q \cdot p_\gamma)} \gamma_\mu P_L+P_R
\gamma_\mu \frac{\not \! p_b -\not \! p_\gamma  +m_b}{(p_b\cdot p_\gamma)}
\not \! \epsilon_\gamma \right] b ~(\bar \nu \gamma ^\mu P_L \nu) ,
\label{h4}
\end{equation}
will be the dominant contributions to the decay $ B_q (b \bar q )
\to \gamma \nu \bar \nu$.

\section{Model calculations }

The effective Hamiltonian given in (\ref{h4})  is not enough to analyze
the processes $B_q \to \gamma \nu \bar \nu $. 
In addition, models are needed to do the 
calculations at the hadronic level.
First we use a simple  constituent quark model
(see, for example \cite{cheng}). 
In this model both of the  (anti)quarks are treated non-relativistically with
the same velocity of the hadron.
 Thus (anti)quarks masses are now constituent quark masses of the order 
of several hundred MeV.
We use further the interpolating field technique\cite{hqet}
which relates all the hadronic matrix elements of the present
case to the decay constants of the mesons.
The amplitude for $B_q \to \gamma \nu \bar \nu $ decay is:
\begin{equation}
{\cal A} = \frac{e C f_{B_q}m_{B_q}}{12( p_{B_q} \cdot p_\gamma)} 
\left[ \left(\frac{1}{m_q}+\frac{1}{m_b}\right) i
 \epsilon_{\alpha \beta \mu\nu}
\epsilon_\gamma^\alpha p_\gamma^\beta p_{B_q}^\nu
+\left(\frac{1}{m_q}-\frac{1}{m_b}\right) 
(p_{\gamma \mu}\epsilon_{\gamma\nu}- p_{\gamma\nu} 
\epsilon_{\gamma\mu})p_{B_q}^\nu \right]
(\bar \nu \gamma ^\mu P_L \nu).\label{5}
\end{equation}
The first term inside the bracket comes from the CP-odd part of $B_q$,
while the second term results from the CP-even part of $B_q$.
Since $m_b>> m_q$ ($q=d,s$), we can safely neglect the term of $1/m_b$.
Then after squaring the amplitude, both CP-odd and CP-even parts
give the same contribution to the decay rate. 
Performing the phase space integration over one of the two Dalitz 
variables, and  summing over
the three generation of neutrinos, we get the 
differential decay width versus the $\nu \bar \nu$ invariant mass:
\begin{equation}
\frac{d\Gamma}{d m_{\nu \bar \nu}^2} =\frac{2 C^2 \alpha f_{B_q}^2 }
{3(48\pi)^2m_{B_q} m_q^2}
(m_{B_q}^2 -m_{\nu \bar \nu}^2) m_{\nu \bar \nu}^2~ .\label{6}
\end{equation}
The decay width is:
\begin{equation}
\Gamma =\frac{C^2 \alpha f_{B_q}^2 m_{B_q}^5 }{(144\pi)^2 m_q^2}
.\label{7}
\end{equation}
Using $\alpha=1/132$, $m_s=0.51$ GeV, 
 $m_t=176$ GeV and $|V_{tb} V_{ts}^*|=0.04$, we get
\begin{equation}
\Gamma (B_s \to \gamma \nu \bar \nu )=9.5\times 10^{-21} \times
\left(\frac{f_{B_s} }{0.2GeV}\right)^2~{\rm GeV}. \label{9}
\end{equation}
If the lifetime is taken as $\tau(B_s)=1.34\times 10^{-12} s$ \cite{pdg},
the branching ratio is found to be $1.9\times 10^{-8}$.
The differential decay rate 
versus  $m_{\nu \bar \nu}^2$
is displayed in Fig.2 (solid line).

For $B_d$ meson decay, we take  $|V_{tb}V_{td}^*|=0.01$ 
and $m_d=0.35$ GeV.  
The decay width is then
\begin{equation}
\Gamma (B_d \to \gamma \nu \bar \nu  )= 1.2\times 10^{-21} 
\times\left(\frac{f_{B_d} }{0.2GeV}\right)^2 ~{\rm GeV},
\end{equation}
which corresponds to the branching ratio  $2.6\times 10^{-9}$, 
if  uses of $\tau(B_d)=1.50\times 10^{-12} s$\cite{pdg} and 
$f_{B_d}=0.2$GeV are made. 

We also apply the pole model to calculate the $B_q$ meson decay.
The main contribution here is the radiative transition of the $B_q$ meson
into an intermediate (virtual) $B_q^*$. 
The Feynman diagram is shown in Fig.3. 
The subsequent process $B_q^* \to \nu \bar \nu$ is determined in SM by 
the effective Hamiltonian   eqn.(\ref{h4}),
while the $B_q B_q^*\gamma$  transition is taken  as
\begin{equation}
{\cal L}=g \epsilon _{\nu \lambda \alpha \beta} \epsilon_\gamma^\nu 
p_\gamma^\lambda p_{B_q}^\alpha \epsilon_{B_q^*}^\beta +{\rm h.c.}.
\end{equation}
This corresponds to the first part of eqn.(\ref{5}), which is 
for the CP-odd part of $B_q$. Hereafter, we will concentrate only on 
the CP-odd combination of $B_q$ and $\bar B_q$.
The effective coupling constant is usually explained as a function of 
the invariant mass of the intermediate $B_q^*$. Here we simply derive
its value from the constituent quark model as a constant \cite{cheng},
\begin{equation}
g= -\frac{e}{3} \left( \frac{1}{m_b} + \frac{1}{m_q}\right).
\label{gcoup}
\end{equation}
The $1/m_b$ term can also be neglected compared to the $1/m_q$ term.
Note that the potential model\cite{potent} gives 
\begin{equation}
g= -\frac{e}{3} \left( \frac{1}{\Lambda_b} + \frac{1}{\Lambda_q}\right),
\end{equation}
with
\begin{equation}
\begin{array}{llr}
\Lambda_d= 0.59 {\rm GeV}, &\Lambda_b= 4.93 {\rm GeV} &~~({\rm for ~~B_d}),\\
\Lambda_s= 0.66 {\rm GeV}, &\Lambda_b= 4.98 {\rm GeV} &~~({\rm for ~~B_s}),
\end{array}
\end{equation}
which is quite close to (\ref{gcoup}).
Now the amplitude of the decay $B_q \to \gamma \nu \bar \nu $ is:
\begin{equation}
A= \frac{C \;g f_{B_q^*} m_{B_q^*} }{\sqrt{2}(m_{B_q^*}^2-m_{\nu \bar
\nu}^2)} \epsilon _{\nu \lambda \alpha \beta} \epsilon_\gamma^\nu 
p_\gamma^\lambda p_{B_q}^\alpha (\bar \nu \gamma ^\mu P_L\nu) .
\end{equation}
Here again we give the differential cross section for the
purpose of being used experimentally:
\begin{equation}
\frac{d\Gamma}{d m_{\nu \bar \nu}^2} =\frac{2C^2 \alpha f_{B_q^*}^2 m_{B_q^*}^2}
{3(48\pi)^2m_{B_q}^3 m_q^2}
\frac{(m_{B_q}^2 -m_{\nu \bar \nu}^2)^3 m_{\nu \bar \nu}^2}{
(m_{B_q^*}^2 -m_{\nu \bar \nu}^2)^2}.\label{13}
\end{equation}
After  integrating over the phase space, the decay width is obtained as 
\begin{equation}
\Gamma =\frac{C^2 \alpha f_{B_q^*}^2 m_{B_q^*}^8 }{(144\pi)^2m_{B_q}^3
m_q^2 }
f(m_{B_q}^2/m_{B_q^*}^2),\label{14}
\end{equation}
where 
$$ f(y)=-5y^3+6y^2+12y-12-6(4-y)(1-y)^2\ln(1-y).$$
The formula (\ref{13},\ref{14}) are similar to the constituent quark
model case eqn.(\ref{6},\ref{7}). It is easy to see that if $m_{B_q^*}=
m_{B_q}$, $f_{B_q^*}=f_{B_q}$ are taken in eqn.(\ref{13},\ref{14}), they
will reduce exactly to eqn.(\ref{6},\ref{7}). 

Now we use $m_{B_s^*}=5.42$ GeV  and get:
\begin{equation}
\Gamma (B_s \to \gamma \nu \bar \nu )=8.8\times 10^{-21} \times 
\left(\frac{f_{B_s^*} }{0.2GeV}\right)^2 ~{\rm GeV}.
\end{equation}
It is only slightly different from eqn.(\ref{9}), because $m_{B_s^*}$ is
not quite different from $m_{B_s}$.
The branching ratio is $1.8\times 10^{-8}$, if $f_{B_s^*}=f_{B_s}=0.2$ 
GeV is used.
The differential decay rate calculated in pole model is also 
displayed in Fig.2 (dashed line) as function of $m_{\nu \bar \nu}^2$.

For $B_d$ meson, we get
\begin{equation}
\Gamma (B_d \to \gamma \nu \bar \nu  )=1.1\times 10^{-21}\times 
\left(\frac{f_{B^*_d} }{0.2GeV}\right)^2~{\rm GeV}. 
\end{equation}
The branching ratio is obtained as $2.4\times 10^{-9}$, also quite close 
to that obtained in constituent quark model.

\section{Conclusion}

We predict the branching ratios in SM for $B_s \to \gamma 
\nu \bar \nu $ to be $10^{-8}$ and for $B_d \to \gamma \nu \bar \nu $ to be
$10^{-9}$. With these branching ratios, they are hopeful to be
 detected at future B factories or LHC. 
They can provide alternate
channels for measuring $f_{B_s}$ ($f_{B_d}$).

\section*{Acknowledgement}

We thank G. Eilam and M. Gronau for helpful discussions.
The research of D.X. Zhang is supported in part by Grant 5421-3-96
from the Ministry of Science and the Arts of Israel.

\newpage
\section*{Figure Captions}
\noindent

Fig.1 Feynman diagrams in standard model for $b\to q \nu \bar \nu$.

Fig.2 Differential decay rate of $B_s \to \gamma \nu \bar \nu$ versus
the $\nu\bar \nu$ invariant mass $m_{\nu\bar \nu}^2 $, with the solid 
line denoting results calculated from constituent quark model, and the 
dashed line corresponding to that of pole model.

Fig.3 Feynman diagram in pole model for $B_q \to \gamma \nu\bar \nu$.

\newpage
\begin{figure}
\centerline{\epsffile{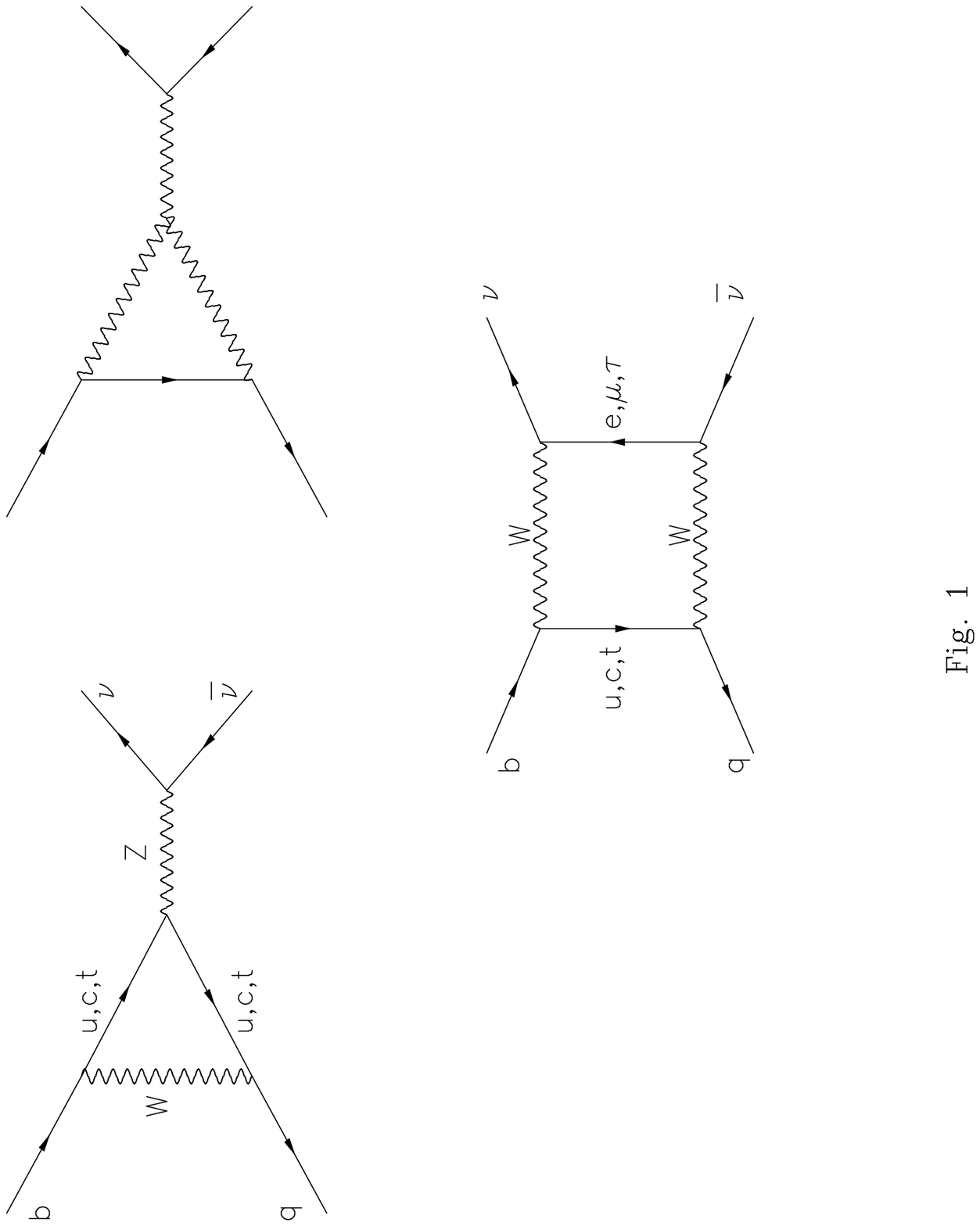}}
\end{figure}

\begin{figure}
\centerline{\epsffile{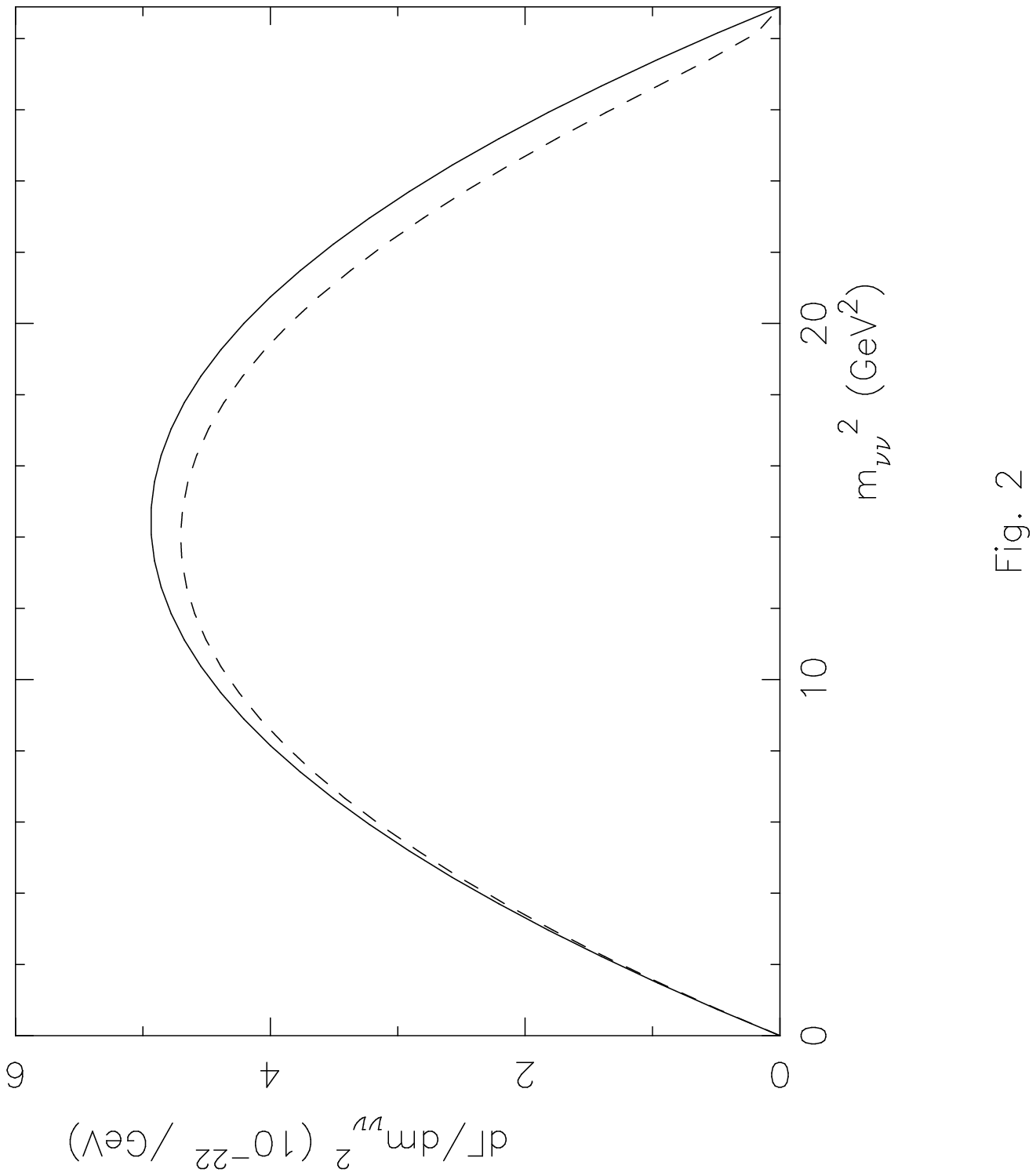}}
\end{figure}

\begin{figure}
\centerline{\epsffile{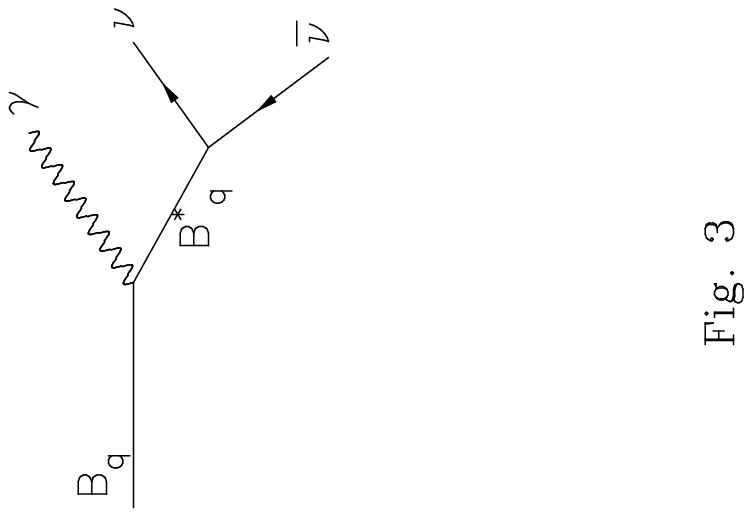}}
\end{figure}

\end{document}